\documentclass[final,preprint,5p,times,twocolumn, authoryear]{elsarticle}

\usepackage{amssymb}

\usepackage{amsmath}

\journal{arXiv.org}

\begin{document}
	
	\begin{frontmatter}
		
		

\title{Regional Disparities and Economic Growth in Ukraine}
\author{Khrystyna Huk}
\author{Ayaz Zeynalov\cortext[*]{\scriptsize Corresponding author: Ayaz Zeynalov, Prague University of Economics and Business, Faculty of International Relations. Address: , 130 67, Prague, Czechia. T: (+420) 224 095 239; \textit{Email:} ayaz.zeynalov@vse.cz. Khrystyna Huk is student at Prague University of Economics and Business, \textit{Email:} hukk00@vse.cz. Financial support from the Prague University of Economics and Business (grant IG212021) is gratefully acknowledged.}}

\affiliation{organization={Prague University of Economics and Business},
	addressline={W. Churchill Sq. 1938/4},
	city={Prague},
	postcode={130 67},
	country={Czechia}}

\begin{abstract}
    This research is devoted to assessing regional economic disparities in Ukraine, where regional economic inequality is a crucial issue the country faces in its medium and long-term development, recently, even in the short term. We analyze the determinants of regional economic growth, mainly industrial and agricultural productions, population, human capital, fertility, migration, and regional government expenditures. Using panel data estimations from 2004 to 2020 for 27 regions of Ukraine, our results show that the gaps between regions in Ukraine have widened last two decades. Natural resource distribution, agricultural and industrial productions, government spending, and migration can explain the disparities. We show that regional government spending is highly concentrated in Kyiv, and the potential of the other regions, especially the Western ones, needs to be used sufficiently. Moreover, despite its historical and economic opportunity, the East region performed little development during the last two decades. The inefficient and inconsistent regional policies played a crucial role in these disparities.
\end{abstract}




\end{frontmatter}

\section{Motivation}
\label{sec:intro}

Regional economic development is one of the pivotal issues faced by developing countries in their medium and long-term development strategies and concerns developed countries. Topic has been central to the plan of economic geography and the more comprehensive social sciences. \citet{Pike2011} claim that deep disparities in the socio-economic development of regions slow down the implementation of effective policy, inhibit the formation of an effective internal market in the country and regions, exacerbate economic processes, and increase social inequality and tension in a society of both developed and developing countries.

The relationship between regional disparities and economic development at the national level is ambiguous. The literature claims that growth as a spatially cumulative process increases regional differences to a great extent \citep{Romer1986, Krugman1991}. The determinants and consequences of regional inequality are the main interest of literature \citep[e.g.,][]{Lessmann2017}. The distributions of natural resources are a major determinant of regional incomes; while some resources are uniformly distributed, others, such as minerals and fuels, are highly concentrated within a country. The agricultural suitability of land is another significant determinant of regional economic development, where arable land contributes to economic growth, dispraising inequality across regions. There is a substantial gap in studying the relationship between regional disparities and economic development in many post-Soviet countries. One of the examples is Ukraine, in which its regions (``oblasts") differ significantly in all dimensions of state functioning, economically, culturally, politically, socially, and in terms of natural resources. We aimed to explain the determinants of regional economic disparities and inequalities in Ukraine.

Although various studies have researched regional development in Ukraine, thus far, quantitative assessment is lacking in the literature. The literature confirms regional disparity in Ukraine and shows that gross regional product growth can explain by export and foreign direct investment in Ukraine in the pre-Crimea conflict \citep{Horska2019}. The authors claim that the role of the resident population on the gross regional product is insufficient, which indicates that miss out attention to regional employment issues and structural imbalance is observed in the labor market at the regional level in Ukraine. \citet{Kallioras2015} highlight two crucial points: the spatial division into Western-Eastern areas is deep with significant deviations in their economic and production structure, and the spatial asymmetry, which is reported down to the regional level, has been corroborated as the regional inequalities have widened. 

\begin{figure}[ht]
	\caption{Gross regional product in Ukraine (2020)\label{fig:1}}
		\includegraphics[width=.48\textwidth]{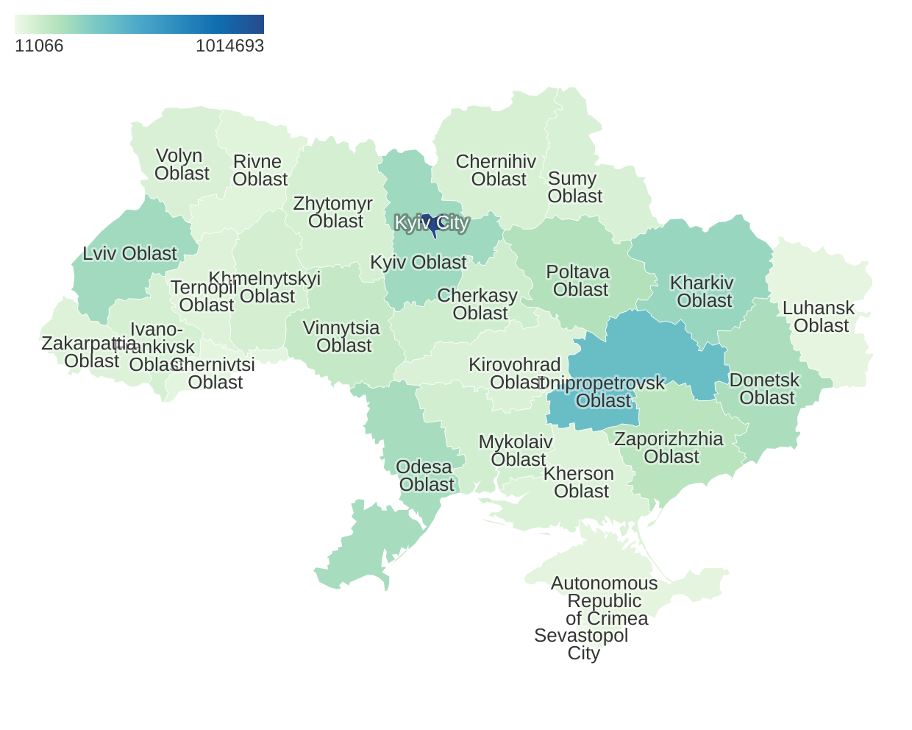}\\
		\includegraphics[width=.48\textwidth]{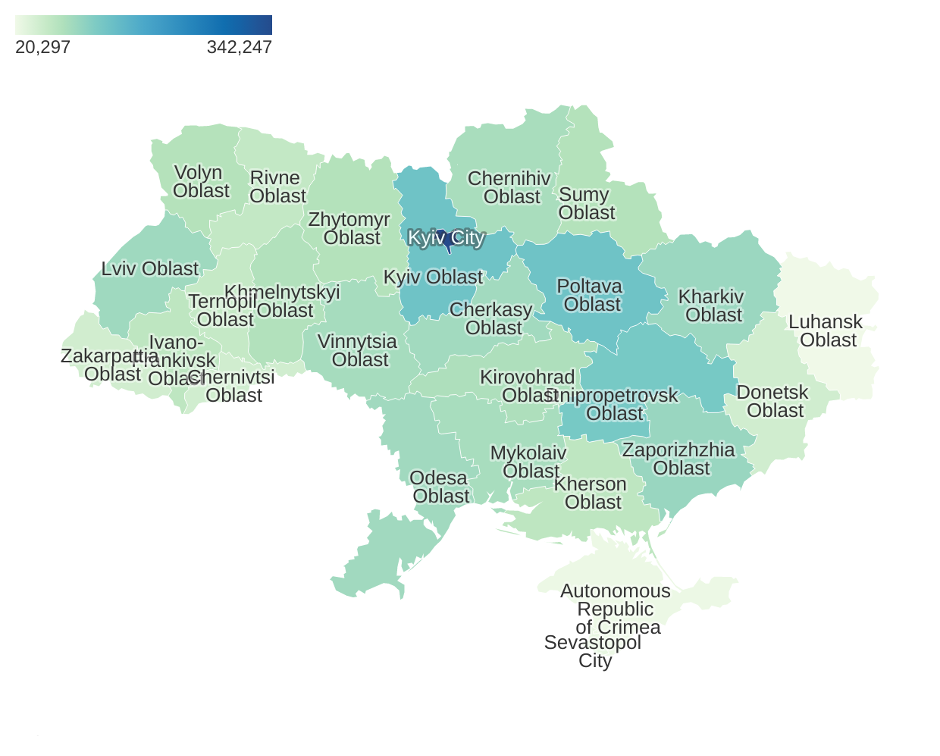}
		\begin{tabular*}{0.45\textwidth}{@{\hskip\tabcolsep\extracolsep\fill}ccccc}
			\multicolumn{4}{p{0.45\textwidth}}{\scriptsize \emph{Note:} The authors' creation is based on data from the State Statistics Service of Ukraine. The above panel presents the Gross regional product with million UAH; the below panel represents Gross regional product per capita with UAH. The last available data for Crimeria is from 2013.}\\ 
		\end{tabular*}
\end{figure}

The challenge of regional disparities in Ukraine relates to economic activities that concentrate on the capital city and two regions. This widen inequality with time as these regions absorb the main share of country investment, technology, knowledge, and human capital. The capital city, Kyiv city, dominates in terms of gross regional product (24 \%), followed by Dnipropetrovsk (10 \%), Kharkiv and Kyiv regions, respectively 6 \% and 5 \% (Figure-\ref{fig:1}). Thus, about 60 \% of the total gross product of Ukraine comes from Kyiv and six regions of Ukraine, which is an indicator of significant regional disparities and uneven development in other regions. For example, the Chernivtsi region lags behind the most, where the share of the gross product in total is about 1 \%. As for the gross regional product per capita, on average, Kyiv city has 3.5 times more than other regions and 7 times more than the regional gross product in the Chernivtsi region.

Regional inequality in Ukraine has deepened during the last decade, especially after the Crimea crisis, and even more during the pandemic period. It is expected to widen more due to the current Ukrainian-Russian war. For instance, \citet{Demchuk2009} do not find significant differences in distributions and aggregate efficiencies between the agricultural and industrial Ukrainian regions, nor between Western and Eastern regions before 2008. However, they found strong support for a rapidly increasing gap between Kyiv city and all the regions since 2001. \citet{Tiffin2006} argues that Ukraine's failure to tap its full potential is mainly a result of its market-unfriendly institutional base. These disparities are crucial for economic development, and improper regional regulation can negatively affect national and regional development. 

The economic and social processes in Ukrainian regions are characterized by considerable unevenness. This, in turn, becomes one of the causes of inequality in the distribution of resources between regions, disturbs the socio-economic balance, and becomes an obstacle to sustainable development. The evolution of the main parameters of regional development and the content of economic relations in economic transformation also demonstrates the growth of disintegration processes. Current conflict will reshape international trade and global value chains, \citet{Estrada2022} emphasize reverse international trade might become more efficient because of current deglobalization and decentralized value chains.

We aim to determine the relationship between regional disparities in industrial and agricultural development, fertility rate, migration, education, capital investments, and regional economic development in Ukraine. The analysis of Ukraine's economic development shows a growing need to ensure the balance of functioning of regional economic systems compared to each other and within the national economic complex. This need is due to both objective factors of the region's development and the lack of experience in implementing public administration, regulatory and economic policies. The spatial heterogeneity of Ukraine's regions complicates the implementation of a unified policy for socio-economic transformation and forming a sustainable national market. It increases the risk of regional crises and interregional conflicts and weakens the integrity of society and the state.

The rest of the paper is as follows: In Section-\ref{sec:background}, we discuss the background of regional disparities in Ukraine. Section-\ref{sec:output} introduces data and research design, presents regional disparity findings. We conclude in Section-\ref{sec:conclusion} by discussing the implications of the findings.

\section{Background on regional disparities}
\label{sec:background}

The reasons for the disparities in the socio-economic development of the Ukrainian regions are various. Historical factors are essential in the differentiation of regions of Ukraine. The affiliation of modern territories to different political formations in the past has influenced various spheres of life and determined the uniqueness of regional development.

\begin{figure}[ht]
	\caption{The Regional Economic Growth Pattern of Ukraine (2004-2020)\label{fig:2}}
	 \includegraphics[width=0.48\textwidth]{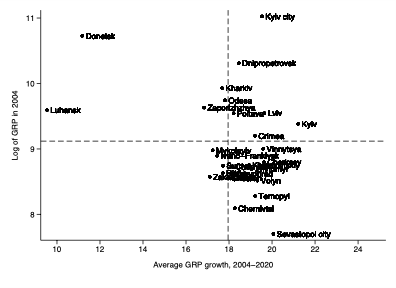}
	 \includegraphics[width=0.48\textwidth]{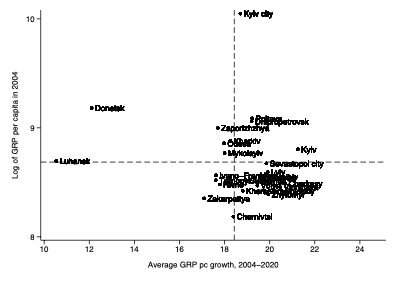}\\
	\begin{tabular*}{0.45\textwidth}{@{\hskip\tabcolsep\extracolsep\fill}ccccc}
	\multicolumn{4}{p{0.45\textwidth}}{\scriptsize \emph{Note:} The authors' creation is based on data from the State Statistics Service of Ukraine. The above panel presents the Gross regional product growth; the below panel represents Gross regional product per capita growth. Lines represent average regional values.}\\ 
	\end{tabular*}
\end{figure}

The transition to a market economy and the establishment of market competition has led to the division of regions according to their competitive advantages and disadvantages, different levels of adaptation to current conditions, and different ability and willingness of local authorities to implement and initiate structural reforms. As a result, Ukraine is currently experiencing deepening regional disparities. Even before the global financial crisis in 2008, researchers drew attention to growing regional inequality in Ukraine. Studying the data for 1990-2007, \citet{Mykhnenko2010} showed that Kyiv, Kharkiv region, and industrial regions of the East concentrated their profits, while the rest of the country either remained at the same level or declined. 

The reasons for the success of these regions are quite different. If such large cities as Kyiv were enriched primarily due to the service and financial economy, Dnipropetrovsk, Donetsk, and Luhansk oblasts would increase their gross product due to industrial production. Other regions have lagged since the agricultural low-productivity output is located mainly there (see Figure-\ref{fig:2}). This situation is closely connected with the model of Ukraine's integration into the global economy and with the strategies of Ukrainian political elites. \citet{Mykhnenko2010} emphasize that until 2004 Ukraine's economic growth was based on creating conditions for developing export-oriented industries, especially metallurgy. This significantly distorted the overall development of Ukraine in the regional context.

The crucial roles in changes in Ukraine’s regional income disparities could be explained by natural and climatic conditions, composition and scale of natural resources, the composition of the population, stage of production and social infrastructure, and level of urbanization. The main share of total coal resources, approximately 92 \%, for example, is in the Donetsk coal basin (Donbass). For comparison, the Lviv-Volyn basin accounts for about 2,5 \% (NISS, 2016).

Poor political and economic decisions may conflict with the social goals of society and public policy that either creates benefits or discriminates against certain regions. For instance, the gap in economic reforms at the regional level, unregulated legal bases, the inefficiency of the implementation mechanism of current legislation, unjustified preferences and benefits provided by the center to individual regions for social support. The lack of a balanced regional policy has led to disparities in the Ukrainian economy's territorial structure, inefficient use of natural resource potential, and research and production potential of the regions.

\begin{figure}[ht]
	\caption{Industrial and Agricultural Production (2020)\label{fig:3}}
	\begin{center}
		\includegraphics[width=70mm]{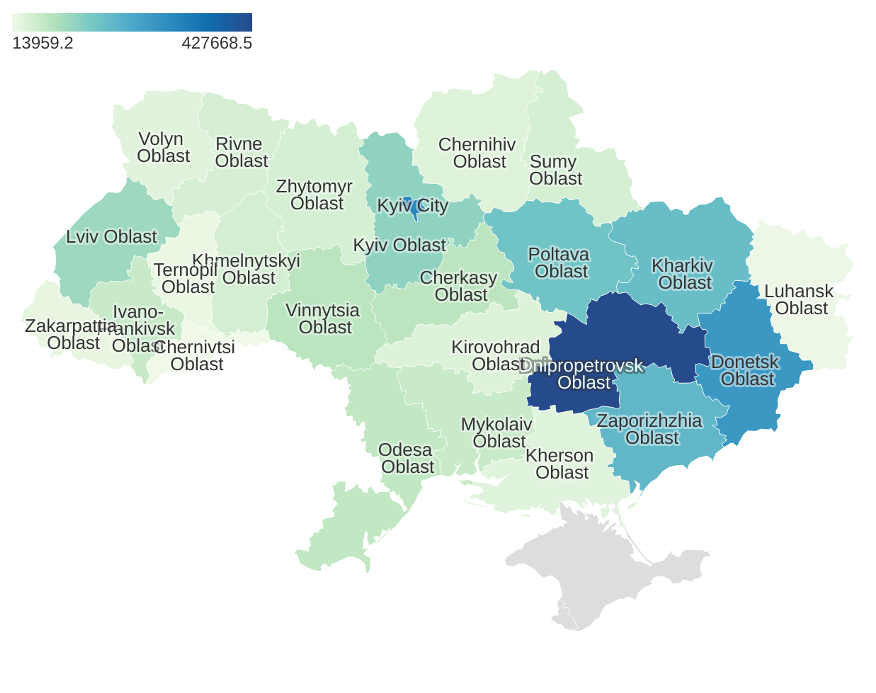}
		\includegraphics[width=70mm]{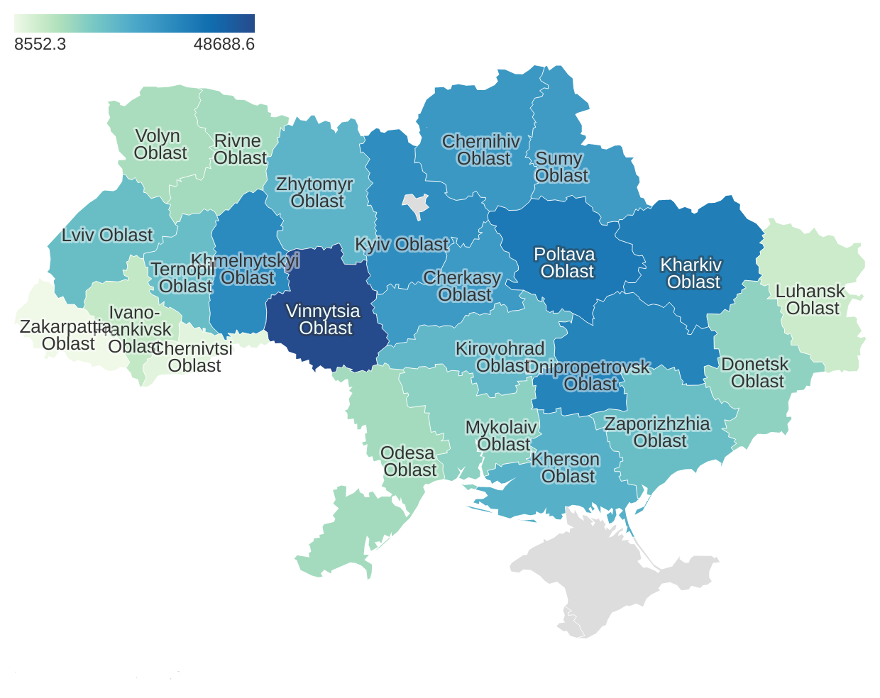}\\
	\end{center}
	\begin{tabular*}{0.45\textwidth}{@{\hskip\tabcolsep\extracolsep\fill}ccccc}
	\multicolumn{4}{p{0.45\textwidth}}{\scriptsize \emph{Note:} The authors' creation is based on data from the State Statistics Service of Ukraine. The left panel presents the industry production volume; the right panel represents agriculture production.}\\ 
	\end{tabular*} 
\end{figure}

The Figure-\ref{fig:3} shows the volume of industrial products sold (goods, services) in Ukraine in 2020 in mln UAH. A significant gap can be clearly seen between Kyiv, Dnipropetrovsk, Donetsk oblasts, and other oblasts. In percentage terms, Kyiv accounts for 13,6 \% of the total value of sold industrial products, and Dnipropetrovsk and Donetsk oblasts 19,1 \% and 15,6, respectively. Half of the sold industrial products and services are carried out in two oblasts and capital city, indicating a significant concentration and unevenness.

\newpage

These are associated with objective factors, such as deposits of natural resources, but also with subjective that is lack of subsidiarity in the relationship between the center and the regional governments. It also relates to low and limited level of self-government; lack of cooperation between oblasts and common criteria for their budget support, which only strengthens regional differentiation. Agriculture production is currently concentrated in Vinnytsia Oblast  (8 \% of total agricultural output), Kharkiv Oblast (6.2 \%), Dnipropetrovsk Oblast (5.9 \%), and Kyiv Oblast (5.6 \%). Agriculture is less concentrated than industry, but there is also uneven development and differentiation of funding for leading agricultural and other areas.

\begin{figure}[ht]
	\caption{Regional capital investments in Ukraine, 2020 (\% of total)}
	\begin{center}
		\includegraphics[width=70mm]{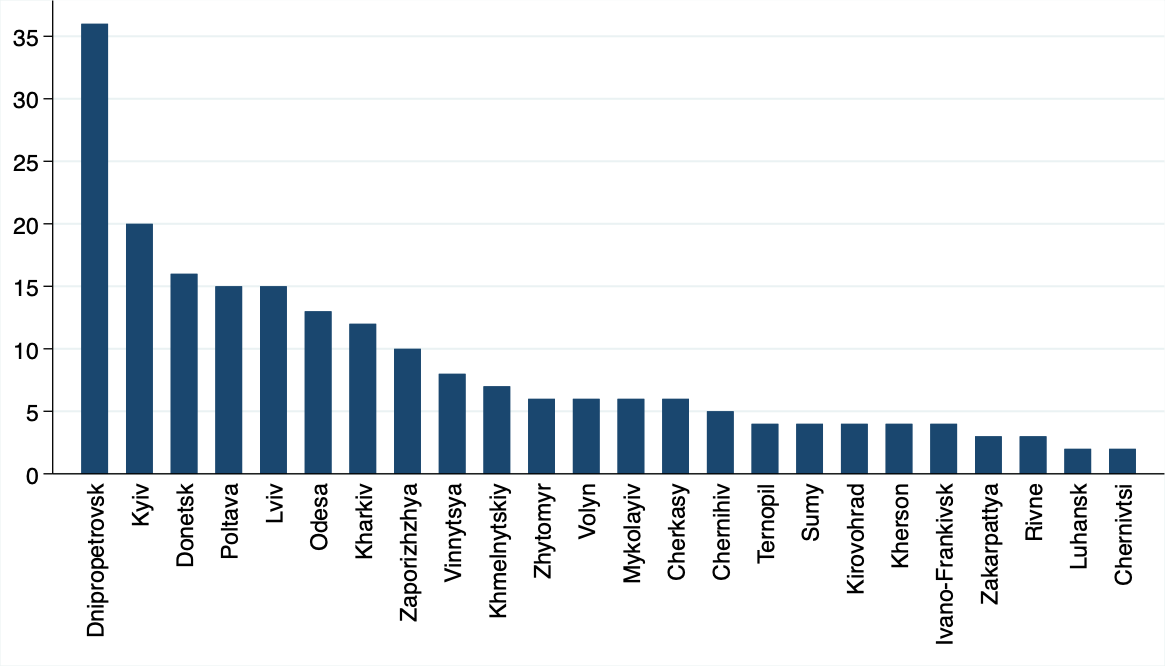}\label{fig:4}
	\end{center}
	\begin{tabular*}{0.45\textwidth}{@{\hskip\tabcolsep\extracolsep\fill}ccccc}
	\multicolumn{4}{p{0.45\textwidth}}{\scriptsize \emph{Source:} State Statistics Service of Ukraine.}\\ 
	\end{tabular*}  
\end{figure}

One of the manifestations of poor political and economic decisions is capital investment distribution (see Figure-\ref{fig:4}). More than 32 \% of total capital investments go to Kyiv, 6.5 \% to the Kyiv region, almost 12 \% to the Dnipropetrovsk region, and 5 \% to partially occupied Donetsk. Every other region accounts for less than 5 \% of total capital investments.

The disproportionate development of the regions, and the emergence of the so-called donor regions, through which other regions developed, is the result of Soviet policy, which dates back to the late 20s of the twentieth century. So-called depressed regions have emerged with corresponding development trends, i.e., declining economic development, increased equipment wear, the outflow of mostly skilled people, declining production efficiency, and deteriorating financial and social situations. According to the Law of Ukraine, a depressed region is ``the region in which the average volume of gross regional product (GRP) (until 2004 – the volume of gross value added) per capita is the lowest over the past five years". Further research on the definition of depressed regions raises the problem of establishing a maximum level of deviation from the average GRP per capita. It is assumed that depressed regions will be determined by the Cabinet of Ministers of Ukraine ``manually."

This approach can be subjective, and the composition of groups in depressed regions can change significantly if the definition of depression and regions is approached from a European position – the limit of depression – is 75 \% of the national GRP per capita.

The first group with a high level of development included highly developed regions – Poltava, Kyiv, and Dnipropetrovsk regions. Quite a high level of development is in Zaporizhia and Kharkiv regions, which form the second group. Odesa, Mykolaiv, Lviv, Cherkasy, Kirovohrad, and Vinnytsya regions have an average level of development and are classified in the third group. The fourth group, below the average level of development, includes Chernihiv and Sumy regions. Quantitatively the largest (9 oblasts) is the fifth group of low-level regions, which includes such depressed regions as Volyn, Donetsk, Zhytomyr, Ivano-Frankivsk, and Rivne, Ternopil, Kherson, Khmelnytsky, and Chernihiv regions. Zakarpattia, Chernivtsi, and Luhansk depressed regions have the lowest GRP development per capita and form the sixth crisis level of the development group. Critical remarks of many scholars on the indicator of depressed regions, i.e., GRP per capita, which characterizes only the region's economic side, encourage the study of regional depression and disproportion based on various indicators.

Summing up, there are two stages in developing interregional disparities in Ukraine. The first, until the mid-90s, was characterized by a sharp separation of regions of raw materials and metallurgical exports. This resulted in the development of disparities, manifested in the designation of regions - leaders and outsiders. Some regions have tried to get closer to the leaders at this stage by exporting agro-industrial products. The second stage began in 1995; interregional disparities manifested themselves in a new capacity and new leaders of separation - Kyiv and Kyiv region. At this stage, the importance of regions with transport, mainly transit functions, has increased.

\section{Research design and output}
\label{sec:output}

In this paper, statistical data for administrative units of Ukraine (oblasts), provided by the State Statistics Service of Ukraine and the Ministry of Finance of Ukraine, is used, namely 24 Ukrainian regions, the Autonomous Republic of Crimea, and two cities with special status, Kyiv, and Sevastopol, for the period 2004-2020. For econometric research, the following socio-economic indicators are considered: regional gross product, regional gross product per capita, industrial and agricultural productions, capital investment, government spendings, population, fertility rate, schooling and migration. Due to the annexation of Crimea and Russia's occupation of the Donetsk and Luhansk regions in 2014, we have discontinuity for some regions.

We use panel data estimation, our baseline model is:
    \begin{equation} 
    Y_{it}= \beta_0 + \alpha X_{it} + \beta^{n}_{s=1} D_i + u_{it}
    \end{equation} 
where $Y_{it}$ represents regional economic growth, $X_{it}$ are the control variables, $D_i$ represents regional dummy variables. While $\alpha$ represents the impact of the selected variable on regional economic growth, $\beta$ represents regional difference.

\begin{figure}[ht]
	\caption{Average Gross Regional Product per capita in Ukraine (1994-2020)\label{fig:5}}
	\begin{center}
		\includegraphics[width=70mm]{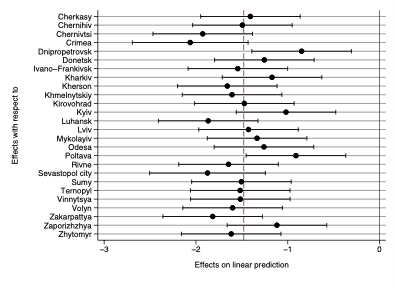}
		\includegraphics[width=70mm]{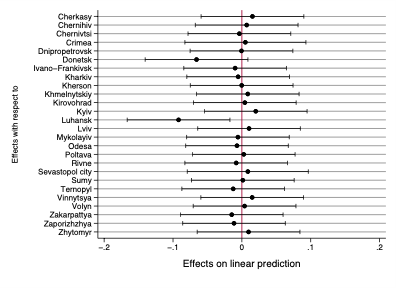}\\
	\end{center}
	\begin{tabular*}{0.45\textwidth}{@{\hskip\tabcolsep\extracolsep\fill}ccccc}
	\multicolumn{4}{p{0.45\textwidth}}{\scriptsize \emph{Note:} The figures represent average marginal effects with 95\% confidence intervals. The above represents average GRP per capita differences comparable to Kyiv city. The dashed red line presents the average value across regions (excluding Kyiv city). The dependent variable is the logarithm of GRP per capita. The below represents GRP per capita growth comparable to Kyiv city.}\\ 
	\end{tabular*} 
\end{figure}

Figure-\ref{fig:5} represents average marginal effects with 95\% confidence intervals. The above represents average GRP per capita differences comparable to Kyiv city. The dashed red line presents the average value across regions (excluding Kyiv city). The dependent variable is the logarithm of GRP per capita. The right represents GRP per capita growth comparable to Kyiv city. We used a random-effect model, controlling regional and year dummy variables. Our results confirm that all regions have lower GRP per capita than Kyiv city (Figure-\ref{fig:5}, left). Dnipropetrovsk, Poltava and Kyiv regions have higher economic output growth. Meanwhile, Crimea, Luhansk Sevastopol, and Zakarpattya performed lower economic output growth between 2004-2020.

Figure-\ref{fig:5} (below) represents GRP per capita growth comparable to Kyiv city. Most regions performed similarly to Kyiv city; however, two regions - Donetsk and Luhansk- have significantly lower economic growth than others, followed by Zakarpattya. Chernivtsi region has a lower GRP per capita.

The one reason of regional disparities in Ukraine can be explained by industrial production. The industrial sector of the national economy forms the financial basis for ensuring sustainable endogenous growth of Ukraine's regions. Despite the slowdown in industrial development in Ukraine due to the influence of many factors (socio-political, monetary), the industry remains the leading type of economic activity.  Our finding is in line with the literature, \citet{Zheng2021} shows that rural industrial development is the driving force for rural economic growth, and foreign direct investment plays a crucial role in regional disparities in China. 
 
The foreign direct investment has weak impacts on industry production due to political instability, weak institutional governance, military conflicts and incomplete reforms in Ukraine \citep{Getzner2020}. \citet{Nguyen2022} claim that  in order to attarct more FDI and sustain able economic growth,  geopolitical stability is essential for emerging country.  The nominal volume of sold industrial products in dollar terms in 2020 was only 69.4 \% of the 2011 level. The nominal volume of products sold in the most technological sectors of the Ukrainian industry (in particular, in mechanical engineering and the chemical industry) from 2011 to 2020 showed a more than two-fold drop. The main reasons for the decrease in the industrial products volume industrial potential of Ukraine, in addition to the ever-increasing competition in the domestic and foreign markets and the uncompetitiveness of the innovative potential of the industrial production, are the difficulties of crediting and weak opportunities for attracting foreign investments, which is associated with changes in power in the country and political instability. 

\begin{figure}[ht]
	\caption{The determinants of regional economic growth in Ukraine (1994-2020)\label{fig:6}}
	\begin{center}
		\includegraphics[width=70mm]{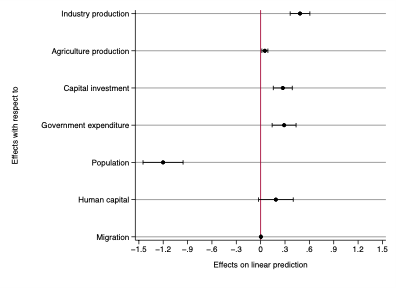}
	\end{center}
	\begin{tabular*}{0.45\textwidth}{@{\hskip\tabcolsep\extracolsep\fill}ccccc}
	\multicolumn{4}{p{0.45\textwidth}}{\scriptsize \emph{Note:} Figure represents average marginal effects with 95\% confidence intervals. Dependent variable is \textit{logarithm of GRP per capita}. Regional and time variables are not reported to save space.}\\ 
	\end{tabular*}
\end{figure}

Next, we assess the determinants of regional economic growth in Ukraine between 2004-2020. Results are presented in Figure--\ref{fig:6}. Industry production has the most significant positive impact on regional economic development. We have only 25 regions since Crimea and Sevastopol city data is unavailable. 1\% increase in industrial production induces a 0.48\% increase in regional economic output. 

The main factors in the fall in the export of industrial goods to the markets of other countries were the annexation of Crimea by Russia, the military confrontation in the East of the country together with the loss of control over part of the territory of Donbas, the loss of cooperation with Russian industrial enterprises, the lack of support for non-raw materials exports from the state, as well as tariff and non-tariff barriers for industrial products of Ukrainian export.

The second reason is that agricultural production significantly impacts economic output, but the effect is ten times lower than industrial production. Agriculture forms food, economic, ecological, and energy security, ensures the development of technologically related industries, and socio-economic foundations of the development of rural areas. The indices of the main indicators of the development of agriculture demonstrate the restoration of the economic potential of agriculture in almost all regions of Ukraine after its significant decrease during 2014-2017. Production volumes increased in 2018 in 22 regions, ranging from 0.5\% in Odesa to 23.7\% in Poltava region. Production volumes also increased significantly in Cherkasy (by 22.5\%), Kyiv (by 20.7\%), Kirovohrad (by 20.4\%), and Sumy (by 11.7\%) regions. As of 2020, the agricultural sector in Ukraine provides an average of 10\% of the GDP and about 40\% of export earnings. Employment of the population in agriculture, fishing, and forestry — 17\% of all employees.

\citet{Fieldsend2019} emphasize that the underdevelopment of the agricultural sector relates to complex knowledge flow between farmers and preference own resources and peer-to-peer knowledge sharing. From 2012 to 2016, agro-food exports declined, but in 2017 they amounted to almost a record 17.8 billion dollars USA. In general, agricultural products and food exports are increasing (up to 50\% by 2020). Monitoring of the agricultural market shows that in recent years this industry has become the flagship of the export of domestic products. According to the results of 2018, agriculture also provided almost 40\% of the country's foreign exchange earnings, surpassing metallurgy by this indicator. Among the problems are the low efficiency of agricultural production in Ukraine compared to other countries, the aggravation of the food problem, the dependence on the import of certain groups of food, the strengthening of structural disproportions in the production of products, the degradation of land resources, the deterioration of the ecology of agricultural areas, the neglect of the development of the social infrastructure of villages, and others. Also, a significant problem is the growth of the level of``agrarianization" in the regions of Ukraine and the deepening of disparities in the regional structure of agricultural production. In 2017, Kherson, Kirovohrad, and Khmelnytskyi regions were the most ``agrarian" in Ukraine (with a share of agriculture in GDP $>$30 \%), while in 2012, there were none.

During 2012-2018, there were noticeable changes in the regional structure of agricultural production. In particular, the shares of 14 oblasts increased, of which Vinnytsia (by 5.6 \%) and Dnipropetrovsk (by 3.9 \%) increased the most. Therefore, these regions (together with Kyiv, Poltava, Kharkiv, Khmelnytskyi, and Cherkasy) became leaders among Ukraine's regions regarding agricultural production. The structure of agricultural production in the regions is dominated by crop production, the share of which in the agricultural production of Ukraine during 2012-2018 increased by a total of 7.2 percentage points. In 2018, the share of this type of agricultural production in the relevant structure exceeded 81.5 \% in 7 oblasts, and it increased most significantly in Cherkasy, Poltava, and Kyiv oblasts. On the other hand, the share of livestock products decreased in all regions over the past six years, remaining the highest in Zakarpattia (48.1 \%) and Ivano-Frankivsk (46.8 \%). The share of animal husbandry decreased most significantly (by more than 10 p.p.) in the Luhansk, Odesa, Mykolaiv, and Dnipropetrovsk regions.

Capital investment and government expenditure are the second most important determinant of regional economic disparities. There is a strong association of capital controls on inflows to mitigate risks to macro stability but not financial stability risks in emerging countries \citep{Das2022}. The role of governance drives regional disparity; insufficient institutional quality leads more significant gap between provinces \citep{Peirpalomino2022}. On the one hand, fiscal decentralisation can lead to more efficient regional development. On the other hand, underdeveloped regions lose competitiveness to better-endowed ones, therefore widening regional disparities gaps \citep{Bartolini2016}.  Subnational government expenditure is constrained and centralised in Ukraine. The ability of local governments to allocate expenditures between and within sectors is quite limited \citep{OECD2018}. The main challenge is  the profitability and transparency of municipal-owned enterprise is low, and it leads non uniform distribution of economic growth.

The disconnection between local communities and community-based organizations is not incentivized in Ukraine \citep{Kvartiuk2019}. The authors claim that people from rural areas trust local governments more than community-based organizations; moreover,  local governments may use community-based organizations as fundraising tools but still keep their autonomy, which leads to inefficient rural development. 

While a higher population reduces economic growth, human capital has substantial positive results. We found that migration does not significantly impact regional economic disparity. The socio-demographic situation in the state is formed depending on the development of population reproduction and migration processes. During the years of Ukraine's independence, the state's population decreased by almost 10 million people. If, in 1991, more than 51.8 million people lived in Ukraine, then at the end of 2020, the general population of Ukraine was 41.6 million people. During 2020, the population of Ukraine, considering migration growth, decreased by 314,062 thousand people, equal to the population of such cities of Ukraine as Sumy, Khmelnytskyi, or Chernivtsi. The tendency to reduce the population of Ukraine has been observed since 1992. Recently, Ukrainian had higher outgoing migration, especially to Poland \citep{Gorny2018}. Poland's agricultural sector dispose of Ukrainian labour migrant workers.

According to the State Statistics Service, in 2020, the volume of population migration growth was the highest in the last four years. The most attractive for internal migration in 2017-2020 was the city of Kyiv, Kyiv, Odesa, Kharkiv, Dnipropetrovsk, and Lviv regions. In all these regions, except for Dnipropetrovsk region, migration growth has been consistently positive over the past four years. On the other hand, Donetsk, Luhansk, Zaporizhzhya, Kirovohrad, Rivne, Vinnytsia, Kherson, Cherkasy, and Chernihiv regions were the least attractive for migration.

A qualitative analysis of the composition of the population of Ukraine in 2017-2020 proved that the internal migration of the population of Ukraine has an urban character; that is, it is associated with the movement of the population from rural areas to cities, and the level of urbanization of the population of Ukraine (the share of the urban population in the total population) is 69.6\%. Information from the State Border Service on the number of trips of Ukrainian citizens abroad for 2017-2020 shows that the most significant number of trips abroad by Ukrainian citizens was to the Republic of Poland - 34,263,750 times or 35.8\% of all trips, Russia - 13,628,262 or 14.2 \%, Hungary - 11,343,768 or 11.9 \% and Moldova - 5,184,105 or 5.4 \%.

The current state of the labor market is complicated by negative demographic processes (increasing mortality and decreasing birth rate due to the migration of young people and young families abroad), which causes negative structural changes in the field of labor resources. The following negative trends characterize the labor market in Ukraine: a decrease in the level of employment, an increase in the number of unemployed, a professional and qualification imbalance and heterogeneity of the situation in the labor market, an ever-increasing number of immigrants and difficulties in their employment following work experience or education. The average level of unemployment in Ukraine in 2005 was 7.2 \% (1.6 million people). In 2006, it decreased to 6.8\% (1.5 million people). In 2007-2008, the unemployment rate was 6.4\%. 

In 2009, the unemployment rate in the country was 8.8\%, and 2 million citizens aged 15 to 70 were without work. In 2010, the unemployment rate was at the level of 8.1\%, in 2011 - 7.9\%, in 2012 - 7.5 \%, in 2013 - 7.2 \%. Since 2014, the unemployment rate in Ukraine began to rise to: 9.3\% in 2014, 9.1\% in 2015, 9.3\% in 2016, and 9.5 \% in 2017. In 2018, 1.6 million economically active citizens, or 8.8 \%, were unemployed. In 2019, the unemployment rate was 8.2 \%, and in 2020 it increased to 9.5 \%. The decline in employment has some consequences. Specific goverment policies are required to insentivize rural development. In China, Targeted Poverty Alleviation program significantly contributed to the earning of people from rural areas \citep{Chang2022}.

According to UNICEF's optimistic scenario, developed based on the latest macro-forecasts of the Cabinet of Ministers, the poverty level in Ukraine may increase from 27.2 \% to 43.6 \% due to a reduction in citizens' incomes. The acute problem is that labor becomes cheaper. The government plans to create jobs with the help of communal and state enterprises through state support and simplifying business conditions. The main shortcoming of these plans is the lack of understanding of how these jobs will stimulate the development of the economy, if they stimulate it at all, and what added value in GDP these workers will generate.

\section{Conclusions}
\label{sec:conclusion}

The national-territorial space of Ukraine today arose because of long-term evolution, when at each new historical stage, new centers of gravity and spatial outlines. One of the most pressing issues in Ukraine today is the existing inequalities in developing individual regions and territories. There are negative trends that have been reflected in the deepening of structural imbalances and resource-reproductive imbalances in the economy of the regions, strengthening interregional socio-economic differentiation of regional development. The presence of a significant regional imbalance of socio-economic development complicates the implementation of a unified policy of economic transformation, the formation of a national market for goods and services, increases the threat of regional crises, and the disintegration of the national economy. Among the areas of current regional policy in Ukraine, gradually forming a promising regional economic and industrial development structure is essential, and ensuring regional justice is a strategic goal and priority of regional development at the national level. 

Currently, the main socio-economic and political capacities are concentrated in Kyiv and other industrial centres. Such regional disparities are primarily the result of the historical development of the state, its regions, and currently imperfect and unbalanced regional policy. We investigated the determinants of regional disparities in Ukraine using panel data estimations. The indicators considered are regional gross product, regional gross product per capita, industry and agricultural productions, capital investment, government expenditure, population, fertility rate, schooling and migration from 2004 to 2020. Our results confirm regional disparities in Ukraine and show what the driving forces are. The first reason is the distribution of natural resources. Industry production concentrated only on specific regions, contributing more significant regional development for those regions. Second is agricultural production, which has a considerable impact but is relatively small. We show that agricultural-rich regions performed poorly compared to industry-based regions. Capital investment and government expenditure play a pivotal role in regional development and are significant contributors to the regional disparities in Ukraine. Incomplete government policies and disconnection between central and regional governments widen regional disparities. 

The reasons behind regional inequality should determine regional policy measures to overcome these interregional disparities. Reducing disparities in the spatial and economic development of regions should be carried out based on government programs that include: the creation of effective mechanisms for implementing regional policy; improving the mechanism of redistribution of resources within the country; creating conditions and a positive image of the regions to attract foreign investment resources; conducting a policy to prevent disintegration processes that threaten the integrity of the country and elimination of factors that lead to exacerbation of social tensions in society.

\bibliographystyle{elsarticle-harv} 
\bibliography{Literature}

\end{document}